\documentclass[10pt]{amsart}
\usepackage[all]{xy}
\usepackage{latexsym}
\usepackage{amsmath}
\usepackage{amsfonts,amssymb}

\newcommand{\beq}{\begin{equation}}
\newcommand{\eeq}{\end{equation}}
\newcommand{\bit}{\begin{itemize}}
\newcommand{\eit}{\end{itemize}}
\newcommand{\ben}{\begin{enumerate}}
\newcommand{\een}{\end{enumerate}}

\begin{document}
\title{A Schematic Model of Gauged S-Duality Spontaneous Breakdown}

\author{Eiji Konishi}

\address{Faculty of Science, Kyoto University, Kyoto 606-8502, Japan}
\email{konishi.eiji@s04.mbox.media.kyoto-u.ac.jp}
\date{\today}
\begin{abstract}We present a schematic model of the $\widehat{sl(2,{\boldsymbol{R}})}$ Kugo-Ojima physical state condition of gauged S-duality using modular forms of the Ramanujan $\Delta$ function type. The properties of the solutions with spontaneously broken gauged S-duality with their dimensionality of space-time and the gauge symmetries of low energy theory are studied.
\end{abstract}
\maketitle
 
 Dynamical gauged S-duality, proposed by the author, has been used to describe the dynamics of D-brane fields.\cite{K,Y1,Y2} The gauged S-duality is defined only by its algebraic structure and, in this formalism, we cannot always evaluate the path-integrals. Nevertheless, we can obtain some non-perturbative properties of the theory by solving the ${\widehat{sl(2,{\boldsymbol{R}})}}$ Kugo-Ojima physical state condition\cite{KO2} of gauged and affinized S-duality.\cite{Konishi2} The search for the solutions of the Kugo-Ojima physical state condition brings us back to Ramanujan's $\Delta$ function.\cite{R} This function is one of the most celebrated modular forms and is defined by
\begin{equation}\Delta[{{q}}]={{q}}\prod_{\ell=1}^\infty\Bigl(1-{{q}}^\ell\Bigr)^{24},\ \ \ {{q}}\in \exp({2\pi i{\mathfrak{h}}})\;,\end{equation}
for the Poincar\'e upper half plane ${\mathfrak{h}}$.

We stress that Ramanujan's $\Delta$ function is a robust prototype of the theory, which, as seen in the following, indicates the $24$ dimensional spatial structure of space-time. This is a very important point of view, especially when we construct string theory by using category theory,\cite{Konishi2} since adopting the category itself in the formulation of the string theory specifies only the formalism, and in the derived category theory there is no robust prototype with a physically nontrivial meaning. The situation of this $\Delta$ function is same as that of the {\it{coordinate frame}} in Euclidean space, which was the most clear expression of the space in classical Newtonian physics and the starting point on its road to the theory of relativity.

This $\Delta$ function results from the discrete modular symmetry originating in the affine Lie symmetry of the model and incorporates the quantization of strings, as seen from the early discoveries in the dual resonance model.\cite{dual} We construct the theory based on this $\Delta$ function by introducing the full $\widehat{sl(2,{\boldsymbol{R}})}$ symmetries, which match the degrees of freedom of the model.

It is worth pointing out that the modularity of this $\Delta$ function is related to the modulus of the world sheet as seen in, for example, the Gliozzi-Scherk-Olive projection in superstring theory\cite{GSO} and the Mandelstam-type open/closed string duality in the partition function of bosonic string theory. The number 24 differs from the critical dimension of bosonic string theory by the dimension of the tachyonic states.

Since the Kugo-Ojima physical state condition on the physical states $\psi$ for the $\widehat{sl(2,{\boldsymbol{R}})}$ Becchi-Rouet-Stora-Tyutin (BRST) charge $Q$ \begin{equation}{{Q}}\psi=0\;,\end{equation}is ${\widehat{sl(2,{\boldsymbol{R}})}}$ covariant, the space of solutions must have $SL(2,{\boldsymbol{Z}})$ modular symmetry.
The dynamics is reduced on the Riemann surface of the moduli space of {\it{affinized part}} of ${\widehat{sl(2,{\boldsymbol{R}})}}$ with an infinite number of time variables due to an infinite number of gauge symmetry generators of $\widehat{sl(2,{\boldsymbol{R}})}$ and Eguchi-Kawai large $N$ reduction.\cite{Konishi2,EK}

Kugo-Ojima physical state condition of the gauged S-duality is infinitesimal description of the moduli space of vacua, so the subject of this note is focused on the {\it{individual}} vacuum in this moduli space. Namely, we will not refer the non-perturbative transitions between different vacuum configurations.

So, we schematically deduce the properties of solutions of Kugo-Ojima physical state condition in terms of modular forms.

Our schematic model of internal gauge symmetries and space-time requires:
\begin{enumerate}
\item The solution in modular form such that the {\it{weight}}, {\it{sum of cusps}} and {\it{index}} of the congruence group are $2n$, $24m$ and $24n$ for integers $n$ and $m$.
The integer $24$ comes from the critical dimension of bosonic string theory minus two. 
\item The {\it{cusps}} of the modular form correspond to the classical saddle-point solution with spontaneous S-duality breakdown. The cusp $i\infty$ corresponds to gravity and is always included. The number of gauge symmetries and their ranks determine the internal space according to the Kaluza-Klein reduction. The standard model demands three kinds of gauge interactions besides gravity. These gauge interactions would be unified in the higher rank terms of the physical state $\psi$. In the modular form, gauge symmetries are encoded as {\it{ideals}} (a notion defined in number theory) in an infinite product expression. An ideal \begin{equation}{{I}}_\ell=\ell{\boldsymbol{Z}}\;,\end{equation} consists of the $\ell$-th saddle point ($\ell=1,2,\cdots,\infty$) of a factor $\chi^{Nn/24}\prod_\ell(1-\bigotimes_{1}^n \chi^\ell)^N$ in the physical state (see Eq.(\ref{eq:solution})). We identify each {\it{prime}} ideal with an independent gauge symmetry. In the analogy with a scalar field with mass $\mu$, i.e., $(\square-\mu^2)\phi=0$, the logarithm of space-time dimension at such saddle points, i.e., $\sum_\ell{\mathrm{ln}}(1-\chi^\ell)=\sum_\ell(-\chi^\ell-\chi^{2\ell}/2-\cdots)$, depends on the {\it{vacuum energy}},\cite{Konishi2} corresponding to the Casimir operator of $\widehat{sl(2,{\boldsymbol{R}})}$.
\item The {\it{genus}} of the gauged S-duality moduli space is desired to be a low one. This requires the axion-dilaton degenerations of $T$-dual type IIB string theory from the original M-theory.\end{enumerate}

The schematic form of the solution when the gauged S-duality is not broken is
\begin{equation}
\psi[\chi]=\chi\prod_{\ell=1}^\infty \Bigl(1-\chi^\ell\Bigr)^{24}\;.
\end{equation}

The variable of the wave function $\psi$ is the physical state $\chi$ corresponding to the Chan-Paton factor of one D-brane which satisfies
\begin{equation}Q_\ell\chi^{ \ell}=0\end{equation}
for the $U(\ell)$ BRST charge $Q_\ell$. 
The wave function $\chi^\ell$, which needs to have no mass dimension, can be expanded in terms of the number of ghosts 
\begin{equation}
\chi^{\ell}=\sum_{\vec{\ell}}\chi_{\vec{\ell}}\prod_{1}^{\ell}{\mathfrak{c}}\;,
\end{equation}
where the coefficients $\chi_{\vec{\ell}}$ are independent of the Grassmann numbers ${\mathfrak{c}}$ of the ghosts and the anti-ghosts.

The degeneracy of vacua in $\psi[\chi]$, that leads to the flat ten dimensionality of space-time, breaks due to the existence of the vacuum energy in the BRST charge.\cite{Konishi2} We call this gauged S-duality spontaneous breakdown (GSSB).

The schematic form of the solution of GSSB that satisfies these requirements is
\begin{eqnarray}
\psi^{(0)}[\chi]=\chi\prod_{\ell=1}^\infty\prod_{n^k\in c_{\Gamma}}\Biggl(1-\biggl[\bigotimes_1^n\chi\biggr]^{\ell}\Biggr)^k\;.\label{eq:solution}
\end{eqnarray}

The form of the solution $\psi^{(0)}$ has the symmetry group $\Gamma$ of subgroup of $SL(2,{\boldsymbol{Z}})$. The cusps of this group $c_{\Gamma}$ are denoted by ${n}^{k}$ with $k$ orbits and width $n$.

\smallskip

We make important remarks about two fundamental structures of these schematic solutions.

\bigskip

\noindent\underline{On the dimensionality of space-time.}

\bigskip

Eq.(\ref{eq:solution}) has the following division of the spatial dimensionality:
 \begin{equation}
 24=\sum_{n^k\in c_\Gamma}(k\cdot n)\;.
 \end{equation}
 
 We remark that GSSB breaks the flat directions of the moduli space of vacua non-perturbatively since the flatness of each direction of space-time is perturbatively stable so that a $\Delta[{{q}}]$ type solution is possible.

We also draw attention to the fact that our formulation is completely background independent, and even the dimensionality is not assumed.

We note that the factor $\chi^{\frac{r}{24}}$ in Eq. (\ref{eq:solution}) arises from the consistency with modularity. The product of Dedekind $\eta$ functions $\prod_{i=1}^k\eta(r_i \chi)$ is a modular form with weight $k/2$ when $\sum_ir_i=24$.

\bigskip

\noindent\underline{On the gauge symmetries.}

\bigskip

However tempting it might be to regard the width of a cusp as the rank of a corresponding gauge symmetry, this will not work. The rank of the gauge symmetry does not appear in each solution explicitly. The gauge symmetries are based on ideals and if one ideal is included by two other ideals, i.e, \begin{equation}{{I}}_1\subset {{I}}_2\cap{{I}}_3\;,\end{equation} the first interaction unifies the other two interactions.

In this note, we can only conclude the number of gauge symmetries and their relations with each other since we may need to solve the Kugo-Ojima physical state condition exactly including some self-consistency conditions to specify the concrete form of the variable $\chi$.

\end{document}